# A comprehensive DFT based insights into the physical properties of tetragonal Mo$_5$PB$_2$

M.I. Naher, M.A. Afzal, S.H. Naqib*

*Department of Physics, University of Rajshahi, Rajshahi 6205, Bangladesh*
*Corresponding author email: salehnaqib@yahoo.com

**Abstract**
Tetragonal Mo$_5$PB$_2$ compound, a recently discovered superconductor, belongs to technologically important class of materials. It is quite surprising to note that a large number of physical properties of Mo$_5$PB$_2$, including elastic properties and their anisotropy, acoustic behavior, electronic (charge density distribution, electron density difference), thermo-physical, bonding characteristics, and optical properties have not been carried out at all. In the present work we have explored all these properties in details for the first time with density functional theory based first-principles method. Mo$_5$PB$_2$ is found to be a mechanically stable, elastically anisotropic compound with ductile character. Moreover, the chemical bonding is interpreted by calculating the electronic energy density of states, electron density distribution, elastic properties and Mulliken bond population analysis. Mo$_5$PB$_2$ has a combination of mainly ionic, metallic, and some covalent bonding characteristics. The compound possesses high level of machinability. The band structure along with a large electronic density of states at the Fermi level reveals metallic character. Calculated values of different thermal parameters of Mo$_5$PB$_2$ are closely related to the elastic properties. The energy dependent optical parameters show close assent to the underlying electronic band structure. The optical absorption and reflectivity spectra and the low energy index of refraction of Mo$_5$PB$_2$ show that the compound holds promise to be used in optoelectronic device sector. Unlike the notable anisotropy found in elastic, mechanical properties and minimum thermal conductivity, the optical parameters are found to be almost isotropic with respect to the polarization direction of the incident electric field.

## 1. Introduction

The binary $T_5M_3$ phases, where $T$ is a transition or rare-earth metal and $M$ a (post)-transition metal or a metalloid element are interesting solids from both application and fundamental physics point of views [1]. This crystal family contains three distinct structural symmetries: tetragonal Cr$_5$B$_3$-type (I4/mcm, No. 140), orthorhombic Yb$_5$Sb$_3$-type (Pnma, No. 62), and hexagonal Mn$_5$Si$_3$- type (P6$_3$/mcm, No. 193). The tetragonal Cr$_5$B$_3$ series contains a wide variety of binary and ternary compounds. One of the interesting subsets of these compounds is the ternary borides with stoichiometries $M_5XB_2$ (where $X$ = P, Si). These systems exhibit many interesting electronic ground states [2]. It has been reported that these ternary borides are formed with many transition metals including the 3$d$ transition metals from V to Co, the 4$d$ transition metals Nb and Mo, and the 5$d$ transition metal W [2]. For instance, $Mn_5SiB_2$ shows ferromagnetic behavior with Curie temperature at room temperature [3]. On the other hand, $Fe_5SiB_2$ and $Fe_5PB_2$ are uniaxial ferromagnets with high Curie temperatures, and have been studied as potential permanent magnet materials [4-6]. Unlike other transition metal silicoborides such as $Fe_5SiB_2$ and $Mn_5SiB_2$ [4, 7], $Co_5SiB_2$ phase shows paramagnetic behavior [8].

Few years back, a new compound, namely $Mo_5PB_2$, was added to the series of the tetragonal $Cr_5B_3$-type. It was synthesized and shown that it exhibits superconductivity (SC) with a critical temperature $T_c$ = 9.2 K [2], which is the highest $T_c$ recorded in this particular family. The experimental and theoretical results simultaneously indicate $Mo_5PB_2$ to be a multiband superconductor like the widely studied high-$T_c$ compound $MgB_2$ [9, 10]. According to electrical resistivity measurements under various applied magnetic fields, the upper critical field of $Mo_5PB_2$ is $\mu_0H_{c2}$ ~1.7 Tesla, which is much higher than that of $Mo_5SiB_2$ (0.6 Tesla) or $W_5SiB_2$ (0.5 Tesla) [11, 12]; the other superconductors in the family.

Besides superconductivity and related parameters, a few of the physical properties, such as structural, electronic band structure, density of states, electrical resistivity, heat capacity, and magnetization of $Mo_5PB_2$, have been studied both theoretically and experimentally so far [1, 2]. Remarkably, most of physical properties, e.g., elastic, bonding (including charge density distribution, electron density difference), acoustic, thermophysical, Mulliken population analysis, and optical properties of $Mo_5PB_2$ have not been explored at all till date. For instance, analysis of the elastic behavior, Cauchy pressure, tetragonal shear modulus, Kleinman parameter, machinability index, macro- and micro-hardness, acoustic velocity (both isotropic and anisotropic), acoustic impedance, anisotropy in elastic moduli, and many more are still unexplored. A number of thermophysical properties, such as melting temperature, thermal expansion, heat capacity, dominant phonon wavelength and minimum thermal conductivity (both isotropic and anisotropic) have not been investigated thoroughly yet. Besides the variation in optical constants with incident photon energy are still unknown. Therefore, it is crucial to get a thorough understanding of the elastic, mechanical, acoustic, electronic, thermal, bonding and optical constants spectra of $Mo_5PB_2$ to unravel its full potential for possible applications. We wish to bridge this significant research gap in this study. This constitutes the primary motivation of the present investigations.

The rest of this manuscript has been arranged as follows: Section 2 consists of a brief description of computational methodology. Section 3 contains the computational results and their analyses. Finally, we have summarized the important features of this work and have drawn pertinent conclusions in Section 4.

## 2. Computational method

CASTEP (CAmbridge Serial Total Energy Package) simulation code [13] has been used for the calculation of ground state properties (structure, elastic, electronic, bonding and optical) of $Mo_5PB_2$ using first principles technique. This code uses quantum mechanical plane wave pseudopotential approach based on the density functional theory (DFT) [14, 15]. The exchange and correlation effects were treated in the framework of Perdewe-Burkee-Ernzerhof (PBE) functional of the Generalized Gradient Approximation (GGA) [16]. Vanderbilt-type ultra-soft pseudopotential [17] has been applied to describe the coulomb potential energy caused by the interaction between the valence electrons and ion cores. Ultra-soft pseudopotential saves massive computational time with little loss of computational accuracy. For Mo atoms $4s^24p^64d^55s^1$, for P atoms $3s^23p^3$, and for B atoms $2s^22p^1$ electrons were explicitly treated as valence electrons to perform pseudo atomic calculations. The geometry optimization of $Mo_5PB_2$, was performed using the Broyden–Fletcher–Goldfarb–Shanno (BFGS) minimization scheme [18]. Convergence of the total energy was achieved by setting the plane-wave cutoff energy to 350 eV. The reciprocal-space integration over the Brillouin zone was carried out using the Monkhorst-Pack scheme [19] with a mesh size of 6 x 6 x 3 $k$-points for $Mo_5PB_2$. Geometry optimization of $Mo_5PB_2$ was performed with the total energy convergence tolerance of $10^{-5}$ eV/atom, maximum lattice point displacement within $10^{-3}$Å, maximum ionic force within 0.03 eVÅ$^{-1}$ and maximum stress tolerance, 0.05 GPa, with finite basis set corrections [20]. An energy smearing width of 0.1 eV has been used. The selected tolerance levels have given reliable estimates of structural, elastic, electronic band structure and optical properties with an optimal computational time. All the first principles calculations of $Mo_5PB_2$ are performed for the ground state with default temperature and pressure of 0 GPa and 0 K, respectively.

The calculated values of single crystal elastic constants $C_{ij}$, of tetragonal crystal structure were obtained using the stress-strain method [21]. Considering symmetry of crystals, a tetragonal system has six independent elastic constants, which are $C_{11}$, $C_{33}$, $C_{44}$, $C_{66}$, $C_{12}$ and $C_{13}$. The calculated values of single crystal elastic constants $C_{ij}$ allowed one to evaluate all the macroscopic elastic moduli, such as the bulk modulus ($B$), Young's modulus ($Y$) and shear modulus ($G$) with the Voigte-Reusse-Hill (VRH) approximation [22, 23].

The optical character of a solid depends on the electronic band structure. It is possible to study the optical constant spectra from the knowledge regarding the complex dielectric function $\varepsilon(\omega) = \varepsilon_1(\omega) + i\varepsilon_2(\omega)$. CASTEP calculates the imaginary part of the dielectric function, $\varepsilon_2(\omega)$, from the momentum matrix elements between the occupied and the unoccupied electronic states employing the formula given below,

$$\varepsilon_2(\omega) = \frac{2e^2\pi}{\Omega\varepsilon_0} \sum_{k,v,c} |\langle \Psi_k^c | \hat{u}.\vec{r} | \Psi_k^v \rangle|^2 \ \delta(E_k^c - E_k^v - E) \qquad (1)$$

where, $\Omega$ is the unit cell volume, $\omega$ is the angular frequency of the incident light, $e$ is the electric charge, the unit vector defines the polarization of the incident electric field, and $\Psi_k^c$ and $\Psi_k^v$ are the conduction and valence band wave functions at a given wave-vector k, respectively. The real part of the dielectric function, $\varepsilon_1(\omega)$, can be determined from the corresponding imaginary part of the dielectric function $\varepsilon_2(\omega)$ via the Kramers-Kronig transformation equation. All the other optical constants, such as the refractive index $n(\omega)$, absorption coefficient $\alpha(\omega)$, energy loss-function $L(\omega)$, reflectivity $R(\omega)$, and optical conductivity $\sigma(\omega)$, can be deduced from the values of $\varepsilon_1(\omega)$ and $\varepsilon_2(\omega)$ [24].

The Mulliken bond population analysis [25] is a widely used method for getting understanding of the bonding characteristics of a material. The projection of the plane-wave states onto a linear combination of atomic orbital (LCAO) basis sets [26, 27] has been used for $Mo_5PB_2$. The Mulliken bond population analysis can be implemented using the Mulliken density operator written on the atomic (or quasi-atomic) basis as follows:

$$P_{\mu\nu}^M(g) = \sum_{g'}\sum_{\nu'} P_{\mu\nu'}(g') S_{\nu'\nu}(g-g') = L^{-1} \sum_k e^{-ikg} (P_k S_k)_{\mu\nu'} \tag{2}$$

and the net charge on an atomic species $A$ is defined as,

$$Q_A = Z_A - \sum_{\mu \in A} P_{\mu\mu}^M(0) \tag{3}$$

where $Z_A$ represents the charge of the nucleus or atomic core.

## 3. Results and analysis
### 3.1 Structural properties
The crystal structure of $Mo_5PB_2$ is tetragonal ($Cr_5B_3$-type) with space group $I4/mcm$ (No. 140). Figure 1 shows the schematic diagram of the crystal structure of $Mo_5PB_2$. The Mo, P and B atoms occupy the following Wyckoff positions in the unit cell [1]: Mo1 at (0, 0, 0), Mo2 at (0.1659, 0.6659, 0.1410), P at (0, 0, ¼), and B at (0.618, 0.118, 0). The unit cell contains twenty Mo atoms, eight B atoms and four $P$ atoms. The geometry optimization of $Mo_5PB_2$ is performed at zero temperature and pressure to disclose the ground state properties. The results of first-principles calculations of structural properties together with their available theoretical and experimental values [1, 2] are presented in Table 1. Our calculated values of lattice constants and volume are in excellent agreement with experimental values [12] indicating high level of reliability of the computational methodology employed.

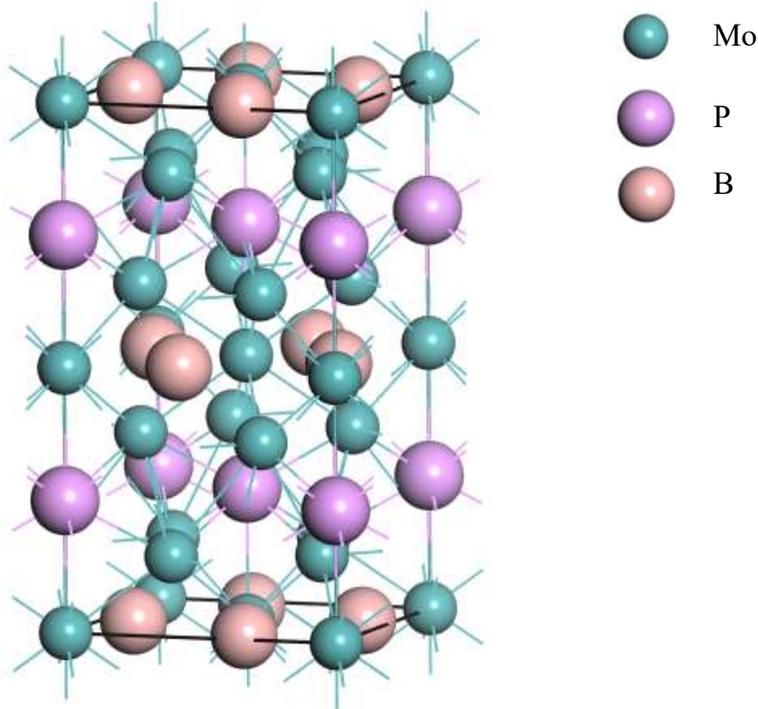

**Figure 1.** 3D schematic tetragonal crystal structure of Mo$_5$PB$_2$ unit cell.

**Table 1.** Calculated and experimental lattice constants $a$ and $c$ (both in Å), equilibrium volume $V_o$ (Å$^3$), total number of atoms in the cell, total number of bonds $N$, and bulk modulus $B$ (GPa) of Mo$_5$PB$_2$.

| Compound | $a$ | $c$ | $c/a$ | $V_o$ | No. of atoms | $N$ | $B$ | Ref. |
|---|---|---|---|---|---|---|---|---|
| Mo$_5$PB$_2$ | 5.97 | 11.07 | 1.85 | 394.67 | 32 | 460 | 277.07 | This |
| | 5.98 | 11.06 | - | - | | | - | [12]$^{Expt.}$ |

### *3.2 Mechanical and elastic properties*

The elastic constants of a crystalline solid provide with the link between mechanical and dynamical behavior of crystal. It also gives important information concerning the nature of chemical bonding in solids. In particular, they give information on mechanical stability (response of a material to applied macroscopic stress), failure modes and stiffness of materials. Therefore, it is important to study elastic constants of a material in details. The calculated values of elastic constants of Mo$_5$PB$_2$ are listed in Table 2. For mechanical stability, according to Born-Huang conditions, a tetragonal system has to satisfy the following inequality criteria [28]: $C_{11} > 0$, $C_{33} >$

0, $C_{44} > 0$, $C_{66} > 0$, $(C_{11} - C_{12}) > 0$, $(C_{11} + C_{33} - 2C_{13}) > 0$, $\{2(C_{11} + C_{12}) + C_{33} + 4C_{13}\} > 0$. All the elastic constants of $Mo_5PB_2$ are positive and satisfy these mechanical stability criteria. This asserts that $Mo_5PB_2$ is mechanically stable with respect to static stresses.

The elastic constants $C_{11}$ and $C_{33}$ give measure of the resistance to linear compressions along [100] and [001] directions, respectively. It is seen that for $Mo_5PB_2$, $C_{11}$ is larger than $C_{33}$, which indicates that the bonding strength/compressibility along [100] directions are stronger/lesser than those along [001] direction in $Mo_5PB_2$. The resistance to shear deformation with respect to a tangential stress applied to the (100) plane in the [010] direction of the compound is represented by the elastic constant $C_{44}$. Here the calculated values show that for $Mo_5PB_2$, $C_{44}$ is much lower than $C_{11}$ and $C_{33}$. This predicts that the compound can be more easily deformed by a shear in comparison to a unidirectional stress along any of the three principal crystallographic directions. The elastic constant $C_{66}$ is related to the resistance to shear of the (100) plane in the [110] direction. For the compound, $C_{44}$ has slightly higher value than $C_{66}$. $C_{11}+C_{12} > C_{33}$ for $Mo_5PB_2$, which predicts that the bonding in the (001) plane is elastically more rigid than that along the $c$-axis and the elastic tensile modulus is higher in the (001) plane than that along the $c$-axis. The dynamical stability of a crystalline material is determined by shear constant, $(C' = \frac{C_{11}-C_{12}}{2})$. It also measures crystal's stiffness (the resistance to shear deformation by a shear stress applied in the (110) plane in the [1$\bar{1}$0] direction). The shear constant is also termed as tetragonal shear modulus. Furthermore, the value of $C'$ is also an indicator of stable and unstable condition of a material. The material is stable if the value of $C'$ is positive, otherwise it is dynamically unstable. The calculated value of $C'$ of $Mo_5PB_2$ is given in Table 2. It is observed that shear constant for $Mo_5PB_2$ is 154.89 GPa (positive) and therefore, $Mo_5PB_2$ is dynamically stable.

The Kleinman parameter ($\zeta$) is another useful parameter, which is an internal strain parameter. The Kleinman parameter is also useful to explain the relative position of the cation and anion sublattices under volume conserving strain distortions for which positions are not fixed by crystal symmetry [29-31]. It is an indicator of the stability of a compound against stretching and bending [29] type of strains. The Kleinman parameter ($\zeta$) of a compound can be estimated using following relation [30]:

$$\zeta = \frac{C_{11} + 8C_{12}}{7C_{11} + 2C_{12}} \qquad (4)$$

It is a dimensionless parameter. The value of $\zeta$ of a compound generally lies between zero to one ($0 \leq \zeta \leq 1$). Where $\zeta = 0$ and $\zeta = 1$ (lower and upper limits) are the indicator of the insignificant contribution of bond bending to resist the external stress and insignificant contribution of bond stretching/contracting to resist the external applied stress, respectively. The calculated value of $\zeta$ of $Mo_5PB_2$ is 0.50 reflecting that mechanical strength in $Mo_5PB_2$ is dominated by both bond bending and bond stretching/contracting contributions.

The Hill values of bulk modulus ($B_H$) and shear modulus ($G_H$) (obtained using the Voigt-Reuss-Hill (VRH) method), Young's modulus ($Y$), Poisson's ratio ($v$) and hardness ($H$) of $Mo_5PB_2$ is evaluated via the following formulae [32-34]:

$$B_H = \frac{B_V + B_R}{2} \tag{5}$$

$$G_H = \frac{G_V + G_R}{2} \tag{6}$$

$$Y = \frac{9BG}{(3B + G)} \tag{7}$$

$$v = \frac{(3B - 2G)}{2(3B + G)} \tag{8}$$

$$H_V = \frac{(1 - 2v)Y}{6(1 + v)} \tag{9}$$

The isotropic (polycrystalline) shear modulus and bulk modulus are well known parameters for the measurement of hardness of a material. The bulk modulus ($B$) is the measurement of resistance to volume change by applied pressure, whereas the shear modulus ($G$) is the measurement of resistance to reversible deformations upon shear stress [35, 36]. Therefore, G is a better predictor of hardness than the $B$. For $Mo_5PB_2$, smaller value of $G$ compared to $B$ (Table 3) predicts that the mechanical strength will be limited by the shear deformation. It is known that the larger value of shear modulus is an indicator of pronounced directional bonding between atoms [37]. A material's covalent nature is manifested from the high values of the bulk and Young moduli [38]. The Young's modulus is defined as the ratio between tensile stress to the tensile strain. The Young's modulus of a material is a measure of the resistance (stiffness) of an elastic material to the change in its length [39, 40] and provides with a measure of thermal shock resistance. A material is stiffer when its value of Young's modulus is large. The calculated Young's modulus is quite high; therefore, $Mo_5PB_2$ is a stiff material. The elastic moduli are linked with a number of thermophysical parameters. For example, the lattice thermal conductivity ($K_L$) and Young's modulus of a compound is related as: $K_L \sim \sqrt{Y}$ [41].

The $G/B$ ratio, which is known as Pugh's ratio, is a widely used parameter to study the brittle/ductile properties of a solid [42-45]. Generally, if the value of $G/B$ is higher/lower than the boundary value of 0.57, the material is brittle/ductile, respectively. In our case, the $G/B$ value of $Mo_5PB_2$ is 0.48 which indicates that the compound is expected show ductile behavior.

The Poisson's ratio, $v$, is one of the crucial parameters for the measurement of compressibility, brittle/ductile nature, and the characteristic of bonding force of a material. The numerical limit for Poisson's ratio of a material is, $-1.0 \leq v \leq 0.5$ [34]. The Poisson's ratio is an indicator of the Ductility-brittleness of solids [46]. A material is brittle, if $v \leq 0.33$; otherwise the material is ductile. The calculated value of Poisson's ratio (0.29) suggests that the compound $Mo_5PB_2$ is ductile in nature. For central forces, the lower and upper limits of the Poisson's ratio of a material are bounded by 0.25 and 0.50, respectively [47, 48]. From the value of ν in Table 3, we can claim that the nature of interatomic forces of $Mo_5PB_2$ is central. The value of Poisson's ratio can also be associated with the presence of ionic and covalent bonding in materials. For ionic and covalent materials, the typical values of ν are 0.25 and 0.10, respectively [49]. The estimated value of Poisson's ratio of $Mo_5PB_2$ predicts that there is significant ionic contribution in the overall chemical bonding in $Mo_5PB_2$.

The Cauchy pressure ($C''$) is another interesting mechanical parameter for solids. The Cauchy pressure of a material is defined as, $C'' = (C_{12} - C_{44})$. A ductile material has positive Cauchy pressure, whereas a brittle material has negative Cauchy pressure [50]. Cauchy pressure is also used to describe the angular characteristics of atomic bonding in a material [51]. Positive and negative values of the Cauchy pressure are associated with the presence of ionic and covalent bonding in a material, respectively. $Mo_5PB_2$ has positive Cauchy pressure, and this clearly indicates that the compound is ductile in nature. The Pettifor's rule [51] states that a material with large positive Cauchy pressure has significant metallic bonds and exhibits high level of ductility. On the other hand, a material with negative Cauchy pressure possesses more angular bonds and thus exhibits more brittleness and significant covalent bonding. Therefore, positive value of Cauchy pressure predicts that some metallic bonding is present in $Mo_5PB_2$ along with appreciable ionic bonding [52].

Machining operations have been a core activity of today's manufacturing industry. One of the main interests of the industry is to achieve either a minimum cost of production or a maximum production rate, or an optimum combination of both, along with better product quality in machining. Thus, machinability index of a material is an important parameter. It is dominated by number of variables, like the elastic properties or characteristics of the work materials, cutting tool material, tool geometry, the nature of tool engagement with the work, cutting conditions, type of cutting, cutting fluid, and machine tool rigidity, hardness and its capacity [53]. The cutting force, tool wear and many more properties of a material are determined by its machinability index. The machinability index, $\mu_M$, of a material can be defined as [54]:

$$\mu_M = \frac{B}{C_{44}} \tag{10}$$

It is also useful for the measurement of plasticity [55-58] and lubricating property of a material. The equation indicates that high tensile strength combined with low shear resistance leads to good machinability and better dry lubricity. A material with large value of $B/C_{44}$ indicates that the corresponding material possesses excellent dry lubricating properties, lower feed forces, lower friction value, and higher plastic strain value. The $B/C_{44}$ value of $Mo_5PB_2$ is 1.91. This predicts that the level of machinability of $Mo_5PB_2$ is quite good; comparable to many technologically prominent MAX phase compounds and alloys [59-63], and other layered ternaries [64-66].

In order to understand elastic and plastic properties of a material, it is also necessary to know its hardness value. The calculated value of Vickers hardness of $Mo_5PB_2$ is 18.42 GPa. This value of Vickers hardness implies that $Mo_5PB_2$ is a fairly hard material [67].

**Table 2**

Calculated elastic constants, $C_{ij}$ (GPa), Cauchy pressure, ($C_{12}$-$C_{44}$) (GPa), tetragonal shear modulus, $C'$ (GPa) and Kleinman parameter ($\zeta$) for $Mo_5PB_2$ at $P = 0$ GPa and $T = 0$ K.

| Compound | $C_{11}$ | $C_{12}$ | $C_{13}$ | $C_{33}$ | $C_{44}$ | $C_{66}$ | $C''$ | $C'$ | $\zeta$ | Ref. |
|---|---|---|---|---|---|---|---|---|---|---|
| $Mo_5PB_2$ | 478.70 | 168.93 | 206.31 | 379.68 | 142.86 | 138.38 | 26.07 | 154.89 | 0.50 | This |

**Table 3**

The calculated isotropic bulk modulus $B$ (GPa), shear modulus $G$ (GPa), Young's modulus $Y$ (GPa), Pugh's indicator $G/B$, Machinability index $B/C_{44}$, Poisson's ratio $v$ and Vickers hardness $H_V$ (GPa) of $Mo_5PB_2$ compound.

| | B | | | G | | | Y | $\frac{G_V}{G_R}$ | $\frac{B_V}{B_R}$ | G/B | $\mu_M$ | v | $H_V$ | Ref. |
|---|---|---|---|---|---|---|---|---|---|---|---|---|---|---|
| $Mo_5PB_2$ | $B_V$ | $B_R$ | $B_H$ | $G_V$ | $G_R$ | $G_H$ | | | | | | | | |
| | 277.80 | 276.35 | 277.07 | 135.19 | 131.40 | 133.30 | 344.63 | 1.03 | 1.01 | 0.48 | 1.91 | 0.29 | 18.42 | This |

## *3.3 Elastic anisotropy*

Since almost all the known crystalline materials are anisotropic with varying degree and a material's direction dependent mechanical behavior can be explained by its elastic anisotropy, it is essential to understand different anisotropy indices. The anisotropy in elasticity influence a large number of important physical processes such as the formation of micro-cracks in solids, motion of cracks, development of plastic deformations in crystals etc. For instance, the shear anisotropic factors give the measure of the degree of anisotropy in the bonding strength for atoms located at different planes. A proper understanding of elastic anisotropy has significant implications in crystal physics as well as in applied engineering sciences. Therefore, it is crucial to estimate elastic anisotropy factors of $Mo_5PB_2$ in details to get information regarding its durability and possible applications under different conditions of external stress.

The degree of anisotropy in the bonding between atoms in different planes can be known by calculating the shear anisotropic factors. The shear anisotropy for tetragonal crystals can be evaluated by three different factors [37, 68]. The shear anisotropic factor for {100} shear planes between the ⟨011⟩ and ⟨010⟩ directions is,

$$A_1 = \frac{4C_{44}}{C_{11} + C_{33} - 2C_{13}} \tag{11}$$

The shear anisotropic factor for the {010} shear plane between ⟨101⟩ and ⟨001⟩ directions is,

$$A_2 = \frac{4C_{55}}{C_{22} + C_{33} - 2C_{23}} \tag{12}$$

and the shear anisotropic factor for the {001} shear planes between ⟨110⟩ and ⟨010⟩ directions is,

$$A_3 = \frac{4C_{66}}{C_{11} + C_{22} - 2C_{12}} \tag{13}$$

The calculated shear anisotropic factors of $Mo_5PB_2$ are enlisted in Table 4. In the case of isotropic crystals, all three factors must be one ($A_1 = A_2 = A_3 = 1$); any other value (lower or higher than unity) implies degree of anisotropy possessed by the crystal. The estimated values of $A_1$, $A_2$ and $A_3$ imply that the compound is moderately anisotropic. $Mo_5PB_2$ shows minimum anisotropy for $A_3$, which is 0.89.

The universal log-Euclidean index is defined as [69, 70],

$$A^L = \sqrt{\left[\ln\left(\frac{B^V}{B^R}\right)\right]^2 + 5\left[\ln\left(\frac{C_{44}^V}{C_{44}^R}\right)\right]^2} \tag{14}$$

where, the Reuss and Voigt values of $C_{44}$ are calculated from [69]:

$$C_{44}^R = \frac{5}{3}\frac{C_{44}(C_{11} - C_{12})}{3(C_{11} - C_{12}) + 4C_{44}} \tag{15}$$

and

$$C_{44}^V = C_{44}^R + \frac{3}{5}\frac{(C_{11} - C_{12} - 2C_{44})^2}{3(C_{11} - C_{12}) + 4C_{44}} \tag{16}$$

The expression for $A^L$ can be used for all crystal symmetries. The universal anisotropy factor, $A^U$ is a relative measure of anisotropy with respect to a limiting value, like other anisotropy indices. It does not give the absolute level of anisotropy. For instance, compounds with higher $A^U$ value do not indicate the presence of higher anisotropy; merely anisotropic nature. This is a reason why the calculation of $A^L$ is sometimes more appropriate, using the distance between the averaged stiffnesses $C^V$ and $C^R$. A crystal is perfectly anisotropic if $A^L$ is equal to zero. The value of $A^L$ for Mo$_5$PB$_2$ is 1.66 which indicates its elastic anisotropy. The values of $A^L$ range between 0 to 10.26, and for almost 90% of the solids $A^L$ greater than 1. Moreover, it has been argued that $A^L$ can also indicate the presence of layered/lamellar type of structure in a solid [35, 53, 69]. Materials with strong layered and non layered structure can be predicted from the higher and lower values of $A^L$, respectively. From the calculated value of Mo$_5$PB$_2$ we can predict that our compound does not exhibit significant layered type of structural configuration, since the value of $A^L$ is comparatively low.

The universal anisotropy index $A^U$, equivalent Zener anisotropy measure $A^{eq}$, anisotropy in compressibility $A^B$ and anisotropy in shear $A^G$ (or $A^C$) for the material with any symmetry can be estimated using following standard equations [69, 71-73]:

$$A^U = 5\frac{G_V}{G_R} + \frac{B_V}{B_R} - 6 \geq 0 \qquad (17)$$

$$A^{eq} = \left(1 + \frac{5}{12}A^U\right) + \sqrt{\left(1 + \frac{5}{12}A^U\right)^2 - 1} \qquad (18)$$

$$A^B = \frac{B_V - B_R}{B_V + B_R} \qquad (19)$$

$$A^G = \frac{G^V - G^R}{2G^H} \qquad (20)$$

Universal anisotropy factor has become one of the widely used anisotropy index due to its simplicity in comparison with the plurality of anisotropy factors defined for specific planes in crystals. Ranganathan and Ostoja-Starzewski [71] have introduced the concept of universal anisotropy index $A^U$, which provides a singular measure of anisotropy irrespective of the crystal symmetry. Unlike all other anisotropy measurement parameters, the influence of the bulk to the anisotropy of a solid was introduced by $A^U$ for the very first time. From Eqn. 17 we can say that a larger fractional difference between the Voigt and Reuss estimated bulk or shear modulus would indicate a stronger degree of anisotropy in crystals. From the values of $G_V/G_R$ and $B_V/B_R$ for Mo$_5$PB$_2$, we can say that $G_V/G_R$ has dominant effect on $A^U$ than that due to $B_V/B_R$. The universal anisotropy factor has only zero or positive value. Whereas the zero value of $A^U$ implies that the crystal is isotropic and any deviation from this value indicates the presence and level of anisotropy. $A^U$ for Mo$_5$PB$_2$ is 0.15. Which is deviated from zero and the compound possesses elastic/mechanical anisotropy.

For an isotropic crystalline material, $A^{eq}$ is equal to 1. The calculated value of $A^{eq}$ for Mo$_5$PB$_2$ is 1.42 predicting that the compound is anisotropic. For an isotropic crystal, $A^G = 0$ and $A^B = 0$. Any deviation from zero (positive) implies the degree of anisotropy. For Mo$_5$PB$_2$, the value of $A^G$ is larger than $A^B$ (Table 4), which indicates that anisotropy in shear is larger than the anisotropy in compressibility.

The linear compressibility of a tetragonal crystal along $a$ and $c$ axis ($\beta_a$ and $\beta_c$) can be evaluated from the following expressions[74]:

$$\beta_a = \frac{C_{33} - C_{13}}{D} \quad \text{and} \quad \beta_c = \frac{C_{11} + C_{12} - 2C_{13}}{D} \tag{21}$$

with $D = (C_{11} + C_{12})C_{33} - 2(C_{13})^2$

The calculated values are listed in Table 4. The calculated values indicate that compressibility along $a$ axis and $c$ axis are anisotropic in nature, whereas anisotropic bulk modulus clearly shows directional anisotropy in Mo$_5$PB$_2$. Compared to all other anisotropy measurement, linear compressibility predicts zero anisotropy in Mo$_5$PB$_2$.

**Table 4**

Shear anisotropic factors ($A_1$, $A_2$ and $A_3$), universal log-Euclidean index $A^L$, the universal anisotropy index $A^U$, equivalent Zener anisotropy measure $A^{eq}$, anisotropy in shear $A_G$ (or $A^C$) and anisotropy in compressibility $A_B$, linear compressibilities ($\beta_a$ and $\beta_c$) (TPa$^{-1}$) and their ratio($\beta_c/\beta_a$) for Mo$_5$PB$_2$ at P = 0 GPa and T = 0 K.

| Compound | $A_1$ | $A_2$ | $A_3$ | $A^L$ | $A^U$ | $A^{eq}$ | $A_G$ | $A_B$ | $\beta_a$ | $\beta_c$ | $\beta_c/\beta_a$ | Layered | Ref. |
|---|---|---|---|---|---|---|---|---|---|---|---|---|---|
| Mo$_5$PB$_2$ | 1.28 | 1.28 | 0.89 | 1.66 | 0.15 | 1.42 | 0.014 | 0.003 | 0.001 | 0.001 | 1 | No | This |

Most of the materials have directional dependent parameters and axial bulk modulus is one of them. With the help of the pressure dependent lattice parameter measurements it is easy to estimate the bulk modulus of a solid along different crystallographic axes. However, it is simpler to estimate the bulk modulus along the crystallographic axes by means of the single crystal elastic constants. The uniaxial bulk modulus along $a$, $b$ and $c$ axis and anisotropies of the bulk modulus can be estimated by using following equations [68]:

$$B_a = a\frac{dP}{da} = \frac{\Lambda}{1 + \alpha + \beta} \quad ; \quad B_b = a\frac{dP}{db} = \frac{B_a}{\alpha} \quad ; \quad B_c = c\frac{dP}{dc} = \frac{B_a}{\beta} \tag{22}$$

and

$$A_{B_a} = \frac{B_a}{B_b} = \alpha \quad ; \quad A_{B_c} = \frac{B_c}{B_b} = \frac{\alpha}{\beta} \tag{23}$$

where, $\Lambda = C_{11} + 2C_{12}\alpha + C_{22}\alpha^2 + 2C_{13}\beta + C_{33}\beta^2 + 2C_{33}\alpha\beta$ and for tetragonal crystals, $\alpha = 1$ and $\beta = \frac{C_{11}+C_{12}-2C_{13}}{C_{33}-C_{13}}$.

where, $A_{B_a}$ and $A_{B_c}$ defines anisotropies of bulk modulus along the *a* axis and *c* axis with respect to *b* axis, respectively.

The calculated values are listed in Table 5. For Mo$_5$PB$_2$, $A_{B_a} = 1$ and $A_{B_b} \neq 1$, which implies anisotropy in axial bulk modulus. Bulk modulus along *c* axis is smaller than those along *a* and *b* axis. Also, the values of uniaxial bulk modulus are different and much larger from the isotropic bulk modulus. This originates from the fact that the pressure in a state of uniaxial strain for a given crystal density generally differs from the pressure in a state of hydrostatic stress at the same density of the solid [37]. These findings are also in good agreement with the result we obtain for linear compressibility.

**Table 5**

Anisotropies in bulk modulus along different crystal axes of Mo$_5$PB$_2$ compound.

| Compound | $B_a$ | $B_b$ | $B_c$ | $A_{B_a}$ | $A_{B_c}$ | Ref. |
|---|---|---|---|---|---|---|
| Mo$_5$PB$_2$ | 1067.36 | 1067.36 | 787.43 | 1 | 0.74 | This |

### *3.4 Acoustic velocities and their anisotropy*

A material's thermal and electrical conductivity depend on another important property, sound velocity. Many of the best room temperature heat conductors (like, diamond and silicon carbide) are crystals with high speed of sound. The study of acoustic behavior of materials have attracted notable interest in physics, materials science, seismology, geology, designing of musical instruments and medical sciences. We have calculated the phase velocity of longitudinal and transverse modes of Mo$_5$PB$_2$. The propagation speed (transverse and longitudinal velocity) of sound in a crystalline material is found from [75]:

$$v_t = \sqrt{\frac{G}{\rho}} \quad \text{and} \quad v_l = \sqrt{\frac{B + 4G/3}{\rho}} \tag{24}$$

where $\rho$ refers to the mass-density of the solid. These equations imply that the sound velocities are strongly determined by density of the compound and transverse velocity does not exist for a material with zero shear modulus.

The average sound velocity $v_a$ can be evaluated using the transverse and longitudinal sound velocities [75]:

$$v_a = \left[\frac{1}{3}\left(\frac{2}{v_t^3} + \frac{1}{v_l^3}\right)\right]^{-\frac{1}{3}} \tag{25}$$

Study of a material having same or different acoustic impedance with the surrounding medium has become a considerable interest in transducer design, noise reduction in aircraft engine, industrial factories, and many underwater acoustic applications [53]. The acoustic impedance is an important parameter which determines the transfer of acoustic energy between two media. When sound is transmitted from one medium (material) to another, the amount of acoustic energy transmitted and reflected at their interface depends on their difference in the acoustic impedance. Thus, if the two impedances are about equal, most of the sound is transmitted. On the other hand, if the impedances differ greatly, most of it is reflected causing transmitted signal loss and echo generation. The acoustic impedance, $Z$, of a material is obtained from [53]:

$$Z = \sqrt{\rho G} \tag{26}$$

where $G$ is the shear modulus and $\rho$ is the density of the material.

The unit of acoustic impedance is the Rayl: 1 Rayl = $kgm^{-2}s^{-1}$ = 1 $Nsm^3$. This equation indicates that a material with high density and high shear modulus has high acoustic impedance.

Another important design parameter in acoustics is the intensity of sound radiation. This parameter is needed for the designing of sound boards, such as the front and back plates of a violin, the sound board of a harpsichord and the panel of a loudspeaker. The radiation intensity, $I$, is proportional to the surface velocity for a given driving source, this scales with modulus of rigidity and density as [53, 76]:

$$I \approx \sqrt{G/\rho^3} \tag{27}$$

where, $\sqrt{G/\rho^3}$ is known as the *radiation factor*. Materials with high values of *radiation factor* are generally used by the instrument makers in order to select materials for suitably designed sound boards. Spruce, widely used for the front plates of violins, has a particularly high value of 8.6 $m^4kg^{-1}s^{-1}$. On the other hand, maple (5.4 $m^4kg^{-1}s^{-1}$) is used for the back plate of violins, the function of which is to reflect musical tones, not radiate [35]. The estimated value of radiation factor for Mo$_5$PB$_2$ is listed in Table 6. In Table 6, we see that longitudinal velocity is almost twice of the transverse velocity.

**Table 6**

Density $\rho$ (g/cm$^3$), transverse velocity $v_t$ (ms$^{-1}$), longitudinal velocity $v_l$ (ms$^{-1}$), average elastic wave velocity $v_a$ (ms$^{-1}$), Acoustic Impedance $Z$ (Rayl) and Radiation factor $\sqrt{G/\rho^3}$ (m$^4$/kg.s) of Mo$_5$PB$_2$.

| Compound | $\rho$ | $v_t$ | $v_l$ | $v_a$ | $Z$ ($\times 10^6$) | $\sqrt{G/\rho^3}$ | Ref. |
|---|---|---|---|---|---|---|---|
| Mo$_5$PB$_2$ | 5.60 | 4879.50 | 9013.06 | 5398.62 | 27.32 | 0.87 | This |

The velocity of propagation of sound (longitudinal and transverse) waves in a material does not depend on frequency and the dimension of the material but only on its nature. There are three modes of vibrations for each atom in a system; one longitudinal and two transverse modes. The anisotropy in sound velocities indicates the presence of elastic anisotropy in a crystal and vice versa. In the case of an anisotropic crystal, the pure longitudinal and transverse modes are only possible along certain crystallographic directions. In all other direction the modes of propagating waves are the quasi-transverse or quasi longitudinal in character. For crystals with tetragonal symmetry, the pure transverse and longitudinal modes can only exist for the symmetry directions of type [010] (or [100]), [001] and [110]. For tetragonal crystal, the acoustic velocities along these principle directions can be expressed as [53]:

*[010] = [100]:*

$[100]v_l = [010]v_l = \sqrt{C_{11}/\rho}; [001]v_{t1} = \sqrt{C_{44}/\rho}; [010]v_{t2} = \sqrt{C_{66}/\rho}$

*[001]:*

$[001]v_l = \sqrt{C_{33}/\rho}; [010]v_{t2} = [100]v_{t1} = \sqrt{C_{66}/\rho}$ \hfill (28)

*[110]:*

$[110]v_l = \sqrt{(C_{11} + C_{12} + 2C_{66})/2\rho}; [001]v_{t1} = \sqrt{C_{44}/\rho}; [1\bar{1}0]v_{t2} = \sqrt{(C_{11} - C_{12})/2\rho}$

where $v_{t1}$ and $v_{t2}$ are the first transverse mode and the second transverse mode, respectively. Directional sound velocities of Mo$_5$PB$_2$ are listed in Table 7.

Both the values of longitudinal and transverse sound velocities are dominated by the values of elastic constants and crystal density of a material. Compounds with lower density and higher elastic constants will have larger velocities of sound. Strong direction dependence in sound velocities is found for Mo$_5$PB$_2$ implying that thermal and charge transport parameters are also expected to be direction dependent.

**Table 7**
Anisotropic sound velocities (in ms$^{-1}$) of Mo$_5$PB$_2$ along different crystallographic directions.

| Propagation directions | | Mo$_5$PB$_2$ |
|---|---|---|
|  | $[100]v_l$ | 9245.66 |
| [100] | $[001]v_{t1}$ | 5050.81 |
|  | $[010]v_{t2}$ | 4970.99 |
|  | $[001]v_l$ | 8234.08 |
| [001] | $[100]v_{t1}$ | 4970.99 |
|  | $[010]v_{t2}$ | 4970.99 |
|  | $[110]v_l$ | 9084.87 |
| [110] | $[1\bar{1}0]v_{t2}$ | 5259.09 |
|  | $[001]v_{t1}$ | 5050.81 |

## *3.5 Thermal properties*

### *3.5.1 Debye temperature*

Debye temperature ($\Theta_D$) is one of the critical parameter controlling thermal properties of solids. Many thermophysical properties, such as thermal conductivity, lattice vibration, interatomic bonding, melting temperature, coefficient of thermal expansion, phonon specific heat and elastic constants of solids depend on the Debye temperature. It also distinguishes between the high- and low- temperature regions of lattice dynamics and heat capacity. Generally, materials with stronger interatomic bonding strength, higher melting temperature, greater hardness, higher mechanical wave velocity and lower average atomic mass exhibit larger Debye temperature. All modes of vibrations possess an energy ~$k_BT$ when the temperature is higher than $\Theta_D$. On the other hand, when $T < \Theta_D$, the higher frequency modes are expected to be frozen and quantum nature of vibrational modes are manifested [77]. The Debye temperature calculated using elastic constants is the same as that determined from specific heat measurements, at low temperatures. This is because at low temperature the vibrational excitation arises solely from acoustic modes (lattice vibration). The Debye temperature of a material, which is proportional to the averaged sound velocity, is derived from [75, 78]:

$$\Theta_D = \frac{h}{k_B}\left(\frac{3n}{4\pi V_0}\right)^{1/3} v_a \qquad (29)$$

where, $h$ is Planck's constant, $k_B$ is the Boltzmann's constant, $V_0$ is the unit cell volume and $n$ is the number of atoms within the unit cell.

The calculated value of the Debye temperature of $Mo_5PB_2$ is listed in Table 8. Debye temperature of $Mo_5PB_2$ is obtained as 429.09 K at 0 K.

### 3.5.2 Melting temperature

Study of the melting temperature ($T_m$) of materials has become an interesting and important area of research. The thermal expansion and bonding energy of a crystalline solid are correlated with its melting temperature. Compounds with higher melting temperature possess strong atomic interaction, higher bonding energy, higher cohesive energy and lower thermal expansion [53]. The melting temperature of materials also gives a rough idea regarding the temperature up to which the materials can be used without substantial oxidation, chemical change, and excessive distortion. The melting temperature $T_m$ of solids can be evaluated, using the elastic constants, from the following equation [79]:

$$T_m = 354K + (4.5K/GPa)\left(\frac{2C_{11} + C_{33}}{3}\right) \pm 300K \qquad (30)$$

The estimated value of melting temperatures of $Mo_5PB_2$ is listed in Table 8. The melting temperature of $Mo_5PB_2$ is 2359.62 ± 300 K which implies that it is a good candidate material for high temperature applications. Compared to many other well known ternaries [67, 80], the melting temperature of $Mo_5PB_2$ is high.

### 3.5.3 Thermal expansion coefficient and Heat capacity

Thermal expansion coefficient ($\alpha$) is an intrinsic thermal property of a material. The thermal expansion coefficients (TECs) and their temperature dependence are of importance in estimating the thermal expansion mismatch with potential substrate materials. It is an important indicator for a materials potential to be used as a thermal barrier coating (TBC). The thermal expansion of materials are correlated with number of other physical properties, namely, thermal conductivity, specific heat, temperature variation of the energy band gap and electron effective mass. It is also important for epitaxial growth of crystals, for the reduction of harmful effects during use in the electronic and spintronic devices. The thermal expansion coefficient of a material can be calculated as follows [53, 76]:

$$\alpha = \frac{1.6 \times 10^{-3}}{G} \qquad (31)$$

where, $G$ is the isothermal share modulus (in GPa). Thermal expansion coefficient of a material is inversely related to its melting temperature: $\alpha \approx 0.02/T_m$ [76, 81]. The thermal expansion coefficient of $Mo_5PB_2$ is tabulated in Table 8.

Heat capacity is a crucial intrinsic thermodynamic parameter of a material. Systems with higher heat capacity exhibit higher thermal conductivity and lower thermal diffusivity. The heat capacity per unit volume ($\rho C_P$) is defined as the change in thermal energy per unit volume in a material per degree Kelvin change in temperature. The heat capacity per unit volume has been calculated using following equation [53, 76]:

$$\rho C_P = \frac{3k_B}{\Omega} \qquad (32)$$

where, $N = 1/\Omega$ is the number of atoms per unit volume. The heat capacity per unit volume of $Mo_5PB_2$ is enlisted in Table 8.

Phonons or quantum of lattice vibrations of materials make a significant contribution to a number of physical properties, such as electrical conductivity, thermopower, thermal conductivity and heat capacity. The dominant phonon wavelength of a compound, $\lambda_{dom}$, is the wavelength at which the phonon distribution function exhibits a peak. The wavelength of the dominant phonon for $Mo_5PB_2$ at 300K has been estimated by using following relationship [35, 81]:

$$\lambda_{dom} = \frac{12.566 v_a}{T} \times 10^{-12} \qquad (33)$$

where, $v_a$ is the average sound velocity in ms$^{-1}$, $T$ is the temperature in degree $K$. Materials with higher average sound velocity, higher shear modulus, lower density exhibit higher wavelength of the dominant phonon mode [35]. The obtained value of $\lambda_{dom}$ (in meter) of $Mo_5PB_2$ is given in Table 8. The calculated value of $\lambda_{dom}$ is 226.12×10$^{-12}$m, which is slightly larger than that for $Nb_2P_5$ (187.78×10$^{-12}$m) a potential thermal barrier coating material [35].

**Table 8**
The Debye temperature $\Theta_D$ (K), thermal expansion coefficient $\alpha$ ($K^{-1}$), wavelength of the dominant phonon at 300K, $\lambda_{dom}$ (m), melting temperature $T_m$ (K) and heat capacity per unit volume $\rho C_P$ (J/m$^3$.K) of $Mo_5PB_2$.

| Compound | $\Theta_D$ | $\alpha$ (×10$^{-5}$) | $\lambda_{dom}$ (× 10$^{-12}$) | $T_m$ | $\rho C_P$ (× 10$^6$) | Ref. |
|---|---|---|---|---|---|---|
| $Mo_5PB_2$ | 429.09 | 1.20 | 226.13 | 2359.62 | 2.10 | This |
| | 501 | - | - | - | - | [12]$^{Expt.}$ |

### 3.5.4 Minimum thermal conductivity and its anisotropy

For high temperature applications of compounds, the behavior of a solid at temperatures above the Debye temperature has become interesting. The thermal conductivity of a compound, at high temperature, approaches to a minimum value known as the minimum thermal conductivity ($k_{min}$). Because of the presence of completely uncoupled phonon modes at high temperature, the heat energy is delivered to neighboring atoms. The minimum thermal conductivity of a compound does not depend on the presence of defects (such as dislocations, individual vacancies and long-range strain fields associated with impurity inclusions and dislocations); which is the prominent feature of it. This is primarily because these defects affect phonon transport over length scales much larger than the inter-atomic spacing. Materials with higher sound velocity and Debye temperature exhibit high minimum thermal conductivity. Clarke deduced the following formula, based on the Debye model, for calculating the minimum thermal conductivity $k_{min}$ of compounds at high temperature [81]:

$$k_{min} = k_B v_a (V_{atomic})^{-2/3} \qquad (34)$$

where, $k_B$ is the Boltzmann constant, $v_a$ is the average sound velocity and $V_{atomic}$ is the total volume per atom [81]. The calculated value of minimum thermal conductivity for $Mo_5PB_2$ is disclosed in Table 9.

Thermal vibrations, movement of free electrons in metals, radiation (if they are transparent) are three ways of transmission of heat solids. Transmission of heat by thermal vibrations involves the propagation of elastic waves. Elastically anisotropic materials naturally exhibit anisotropy in minimum thermal conductivity. The anisotropy in minimum thermal conductivity depends on the longitudinal and transverse sound velocity along different crystallographic directions. The minimum thermal conductivity along different directions, using Cahill model, can be calculated from the following expression [82]:

$$k_{min} = \frac{k_B}{2.48} n^{2/3} (v_l + v_{t1} + v_{t2}) \qquad (35)$$

and $\qquad n = N/V$

where, $k_B$ is the Boltzmann constant, $n$ is the number of atoms per unit volume and $N$ is total number of atoms in the cell with volume $V$.

The minimum thermal conductivity of $Mo_5PB_2$ along [100], [001] and [110] directions are shown in Table 9. Minimum thermal conductivity along [001] is slightly lower compared with those along the other two directions.

To compare results obtained from Cahill's and Clarke's model, we have calculated minimum thermal conductivity of $Mo_5PB_2$ using both the models. The calculated results are listed in Table 9.

The $k_{min}$ of $Mo_5PB_2$ using Cahill's and Clarke's models are 1.46 and 1.04, respectively. We have also seen from previous study that Clarke model predicts lower minimum thermal conductivity [35].

**Table 9**
The number of atoms per mole of the compound, $n$ (m$^{-3}$) and minimum thermal conductivity (W/m.K) of $Mo_5PB_2$ along different crystallographic directions evaluated by Cahill's method.

| Compound | $n$ (10$^{28}$) | $[100]k_{calc.}^{Min}$ | $[001]k_{calc.}^{Min}$ | $[110]k_{calc.}^{Min}$ | $k_{min}$ | | Ref. |
|---|---|---|---|---|---|---|---|
| | | | | | Cahill | Clarke | |
| $Mo_5PB_2$ | 5.068 | 1.50 | 1.42 | 1.51 | 1.46 | 1.04 | This |

## *3.6 Electronic properties*
### *3.6.1 Band structure*
The study of the electronic band structure is one of the most important topics in solid state physics which provides with a complete mapping of energy dispersions of electrons in a crystal structure. Electrical conductivity, Hall effect, thermopower, bonding properties, optoelectronic behavior, superconductivity and itinerant magnetic order, etc. depends strongly on the electronic band structure of materials. The electronic energy band structure of $Mo_5PB_2$, as a function of energy $E$, along different high symmetry directions ($Z$-$A$-$M$-$\Gamma$-$Z$-$R$-$X$-$\Gamma$) in the first BZ has been computed at zero pressure and temperature and is shown in Fig. 2. The Fermi level ($E_F$) is represented by the horizontal broken line placed at zero energy. The total number of energy bands for $Mo_5PB_2$ is 334.

From Fig. 2 it is seen clearly that conduction band and valence band overlap, so there is no band gap at the Fermi level. This indicates that $Mo_5PB_2$ is metallic in nature. Bands crossing the Fermi level are shown in colored lines. These bands exhibit both electron- and hole-like features. The energy bands around the Fermi level mainly come from the Mo $4d$ electronic states. This indicates that the electrons in the Mo $4d$ states mainly dominate the charge conduction in $Mo_5PB_2$.

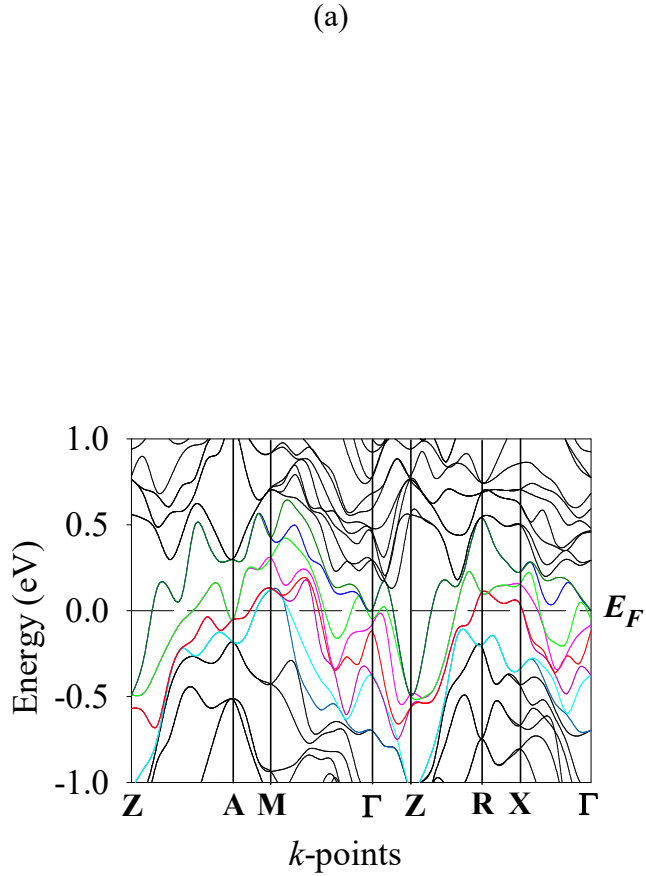
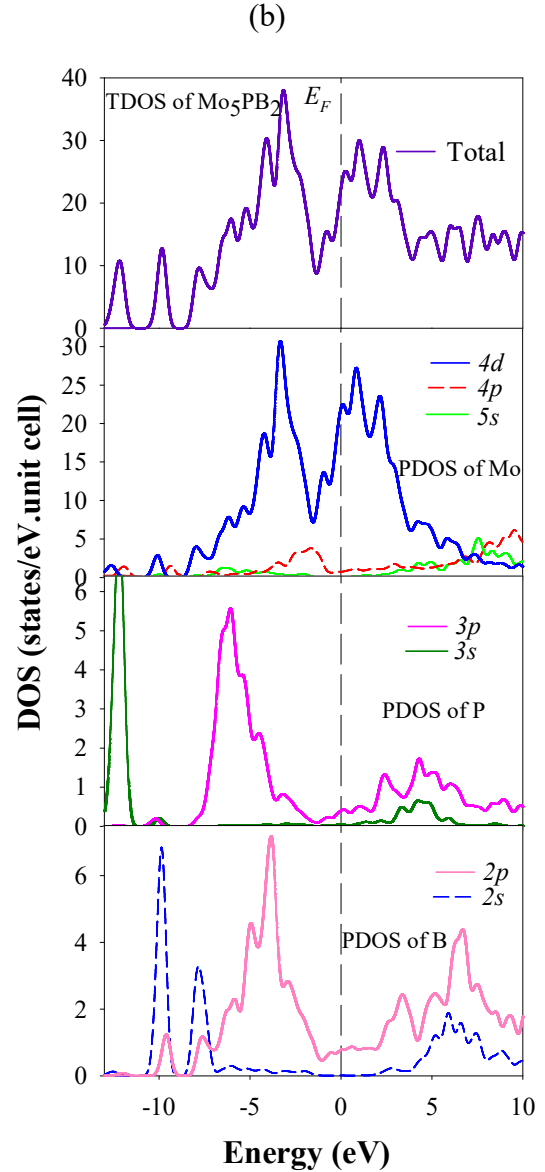

Figure 2. Electronic band structure of Mo$_5$PB$_2$ along several high symmetry directions in the first BZ.

Figure 3. Total and partial electronic density of states (PDOSs and TDOSs, respectively) of Mo$_5$PB$_2$ as a function of energy. The Fermi level is placed at zero energy.

### 3.6.2 Density of states (DOS)

In condensed matter physics, the electronic energy density of states (DOS) of a system defines the number of electronic states available to be occupied per unit energy interval. A material's DOS is a very important parameter to understand the contribution of each atom to bonding, conductivity, magnetic order, optoelectronic properties and many more. Figure 3 represents the calculated total and partial density of states (TDOS and PDOS, respectively) of Mo$_5$PB$_2$ above and below the Fermi level at zero pressure and temperature. The vertical broken line represents the Fermi level, $E_F$. The non-zero value of TDOS at the Fermi level implies that $Mo_5PB_2$ will exhibit metallic electrical conductivity. We have calculated the orbital resolved PDOS of Mo, P and B atoms to understand their contribution to the TDOS. The dominant contribution to the TDOS in the vicinity of $E_F$ comes mainly from Mo 4$d$ state. At the Fermi level the value of Mo 4$d$ is 22.45 states per eV per unit cell. TDOS value of $Mo_5PB_2$ at the Fermi level is 23.73 states per eV per unit cell. Contribution from Mo 4$d$ states amounts to 94% of the TDOS. Thus, the electrical conductivity, chemical and mechanical stability of the compound under study is dominated by the Mo 4$d$ electronic states. The TDOS of $Mo_5PB_2$ at the Fermi level is quite high. This implies that high electrical conductivity and electronic thermal conductivity are expected. The electronic heat capacity is also expected to be high. There are large peaks in the TDOS close to the $E_F$ located at -3.16 and 1.0 eV. These bonding/anti-bonding peaks are formed by the hybridization between Mo 4$d$ orbitals and B 2$p$ orbitals. The response of $Mo_5PB_2$ to the elastic distortion, doping, temperature, etc. are therefore, predicted to be affected by these electronic orbitals. The PDOS close to the Fermi level due to the B 2$p$ orbitals is low; this again implies the major role of the Mo 4d in controlling physical properties of $Mo_5PB_2$.

The computed PDOS can be divided into three energy regions: the lowest energy region is due to the P 3$p$, B 2$s$ and B 2$p$ electronic states; the region from -5 eV to 5 eV are stemming mainly from the Mo 4$d$, P 3$d$ and B 2$p$ electronic states; the region above this is mainly due to the B 2$p$, B 2$s$, P 2$p$ electronic orbitals.

The position of Fermi level and the value of the TDOS of a material, at the Fermi energy, $N(E_F)$, is related to its electronic and structural stability [83, 84]. Pseudogap or quasi-gap is defined as the presence of deep valley in the DOS curve in the vicinity of the Fermi level. This gap is the separation between bonding states and nonbonding/antibonding electronic states. It is related to the electronic stability of materials [85, 86]. For $Mo_5PB_2$, Fermi level falls in the antibonding region (see Figs. 3). The origin of the pseudogap or the quasi-gap usually originates because of two mechanisms [85, 87]. One is due to charge transfer (ionic origin), while the other one is due to hybridization among atomic orbitals. The presence of the directional bonding can be explained by the pseudogap around the Fermi level [88] which facilates covalent bonding and enhances the mechanical strength of material. To be specific, around the Fermi level, a bonding hybridization is seen from the Mo 4$d$ states with B 2$p$ states, which thus exhibits a tendency towards formation of directional covalent bonding between Mo and B. This proposition agrees with the electronic charge density (distribution and difference) mapping and Mulliken bond population analysis results (presented in Section 3.6.3 and 3.7). Besides structural stability, density of states of a material gives much other useful information. (1) Firstly a material has larger ordering energy and high melting point, if the Fermi level lies exactly at the pseudogap. (2) If the Fermi level lies to the right of the pseudogap (in the antibonding region), which refers to an energetically unfavorable state, the system exhibits a tendency to be in the disordered or glassy state [89]. Since, $N(E_F)$ of $Mo_5PB_2$ has high value in the antibonding states, the system can exhibit fairly high superconducting transition temperature, high Pauli paramagnetic susceptibility and electronic specific heat coefficient [90]. (3) If the Fermi level lies in the higher energy antibonding region, the material may show tendency towards electronic instability and the system may try to go to some other configuration to minimize the energy. (4) On the other hand, if the Fermi level lies to the left of the pseudogap (in the bonding states) and all bonding states are not completely filled, additional electrons can make the structure more stable [90]. The interaction of charges among bonding atoms is very crucial for a material's stability; materials possessing higher number of bonding electrons are structurally more stable [91, 92].

The electron-electron interaction parameter of a material, known as the Coulomb pseudopotential, can be estimated from the following relation [93]:

$$\mu^* = \frac{0.26 N(E_F)}{1 + N(E_F)} \tag{36}$$

The total density of states at the Fermi level for $Mo_5PB_2$ is 23.73 states/eV.unit cell. The electron-electron interaction parameter of $Mo_5PB_2$ is therefore found to be 0.25. This value is quite high. The repulsive Coulomb pseudopotential is responsible for the reduction of the transition temperature, $T_c$ of superconducting compounds [89, 93, 94].

### *3.6.2 Electronic charge density distribution*

To illustrate the bonding structure and charge transfer among the atoms in $Mo_5PB_2$, we have calculated electronic charge density distribution around the atoms within the crystal. The electronic charge density (e/Å$^3$) distribution of $Mo_5PB_2$ had been calculated in different crystal planes. Figs 4 show the electronic charge density distribution of $Mo_5PB_2$ in the (100), (001) and (111) planes. The color scale on the right hand side of the charge density maps illustrates the total electron density. The red color indicates low charge (electron) density, whereas the blue color indicates high charge (electron) density. The overall charge distribution agrees reasonably well with the Mulliken bond population analysis (Section 3.7). For (001) plane, the charge density around B atoms is negligible compared to the charge density around Mo atoms. For plane (100), the charge density distribution centered on the Mo atoms are much larger than that around the P atoms. On the other hand, the charge density distribution in (111) plane shows different result (Fig. 4c). It can be seen that electron density is enhanced around the Mo atoms compared to the others (Mo, P atoms). Almost spherical charge accumulation with different signs around the atomic species indicates significant ionic contribution to the overall chemical bonding in $Mo_5PB_2$. There is smearing of background charge in this compound which implies that metallic bonding also contributes. Some deviations from the spherical charge distribution is found for the B atoms which point towards covalent bonding between these atoms. Charge density distribution of $Mo_5PB_2$ on (100) and (001) planes are different, which shows both direction and plane dependency. Therefore, $Mo_5PB_2$ has anisotropic charge density distribution.

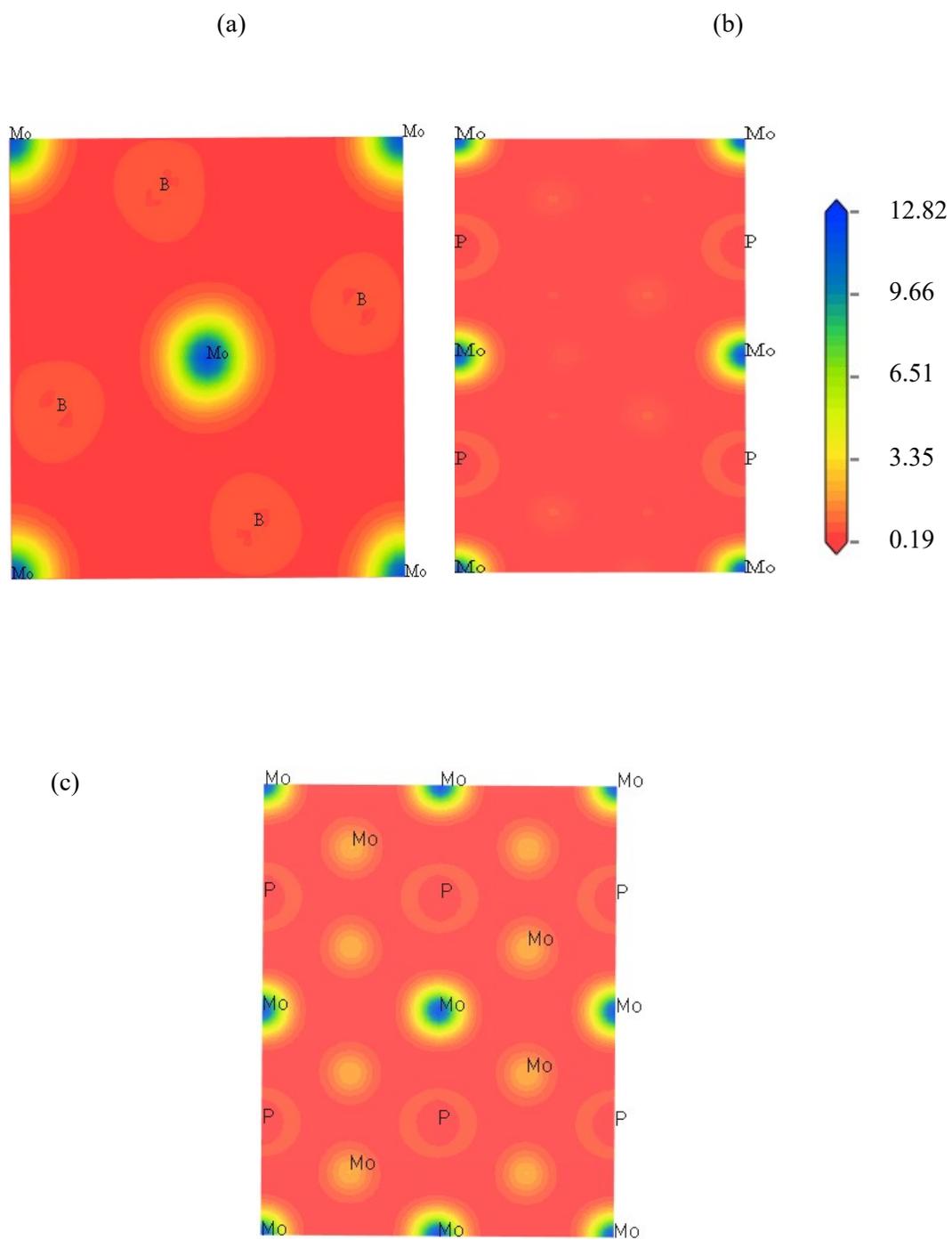

**Figure 4.** Charge density distribution in various crystal planes [(a) (100), (b) (001), and (111)] of $Mo_5PB_2$. The charge density scale is shown on the right.

## 3.6.3 Electron density difference

The electron density difference is another useful means to visualize bonding characters and charge transfer within compounds. Figs. 5(a) and 5(b) show the electron density difference in $Mo_5PB_2$ in (101) and (111) planes, respectively. The color scale shown on the right hand side of charge density difference maps of $Mo_5PB_2$ illustrates the electron density difference. The scale of this map (32 contour lines) is from -0.300 to 0.105 electrons/Å$^3$. The charge density difference maps complement the findings from the electronic charge density distribution analysis very well.

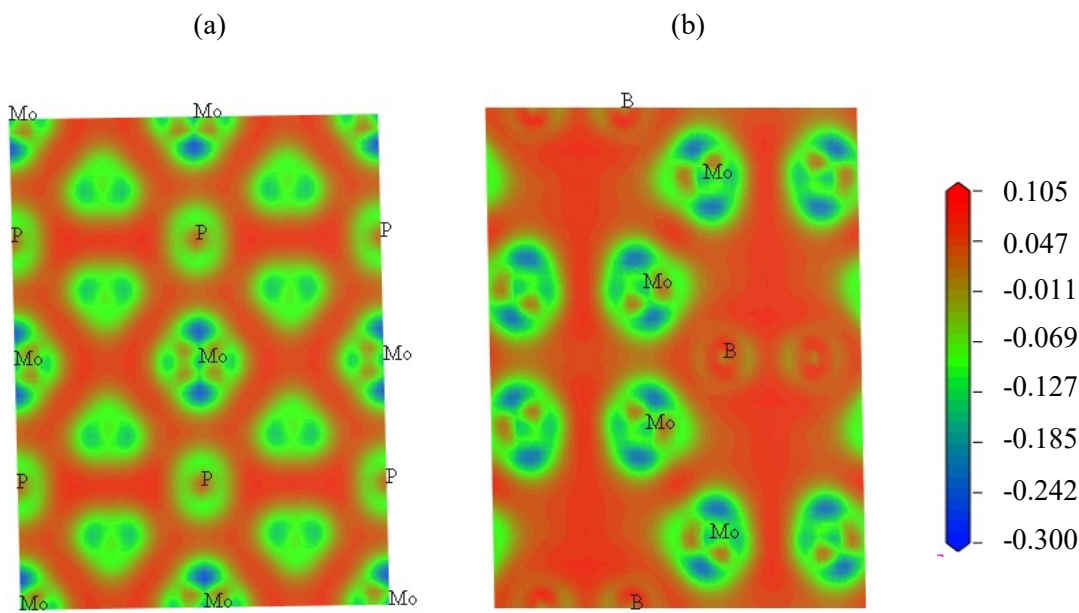

**Figure 5.** The electron density difference map of $Mo_5PB_2$ in (a) (111) and (b) (101) planes.

## 3.7 Bond population analysis

To get more insightful information on the bonding nature (ionic, covalent and metallic) in a compound, and effective valence of atoms in the molecules present in the compound, we have studied the widely used method known as the Mulliken bond population analysis [25]. The Mulliken charge is associated with the vibrational properties of molecules and quantifies how the electronic structure changes under atomic displacement. Dipole moment, polarizability, electronic structure, charge mobility in reactions and other related properties of molecular systems [53] are also related to Mulliken charge. The spilling parameter and atomic charges resulting from these calculations are listed in Table 10. The analysis shows that the total charge for Mo atoms is much larger than P and B atoms. This mainly comes from 4$p$ and 4$d$ states of Mo atom. From Table 10, it is observed that the charge of Mo atoms is both positive and negative (depending on the site within the crystal), which indicates that Mo atoms behave as cation and anion. The atomic charges of Mo in $Mo_5PB_2$ are found to be -0.08 and 0.22 electron.

Whereas the atomic charge of P and B in $Mo_5PB_2$ are 0.12 and -0.47 electron, respectively. All of these charges are deviated from their formal charge expected for purely ionic state (Mo: +4, P: -3 and B: -3). Therefore, there are charge transfers among the atoms. This deviation from formal ionic charge reflects the presence of covalent bonds with ionic contribution.

The detailed effective ionic valence is also presented in Table 10. The effective valence for an atom is defined as the difference between the formal ionic charge and the Mulliken charge [27]. It is useful for measuring the degree of covalency and/or iconicity. We have calculated the effective valence of all the atoms in $Mo_5PB_2$. The zero value of effective valence is associated with a perfect ionic bond, whereas values greater than zero indicate increasing level of covalency. Table 10 shows that the effective valences for Mo is higher compared to P and B in $Mo_5PB_2$. Which indicates the presence of both ionic and covalent bonds are present in $Mo_5PB_2$.

Since early days, it has been found that the Mulliken population analysis (MPA) display extreme sensitivity to the selected atomic basis set and due to this strong basis set dependence; sometimes it produces results contradictory to chemical intuition. On the contrary, Hirshfeld population analysis (HPA) [95] does not require a reference to basis set or their respective location, thus it gives more meaningful result. Taking this into account, we have calculated Hirshfeld charge of $Mo_5PB_2$ using the HPA. For $Mo_5PB_2$, HPA shows different result compared to Mulliken charge. Hirshfeld analysis shows atomic charge of Mo are +0.12 and +0.16 electronic charge, whereas it shows atomic charge of P and B are -0.20 and -0.27 electronic charge, respectively. Hirshfeld charge predicts that electrons are transferred from Mo to P and B. Both the approaches predict the presence of covalent and ionic bondings between Mo, P and B atoms. We have also estimated effective valences of $Mo_5PB_2$ from the Hirshfeld charge. Therefore, HPA predicts almost same level of covalency of $Mo_5PB_2$ compared to the result we got from the Mulliken charge analysis.

**Table 10**

Charge spilling parameter (%), orbital charges (electron), atomic Mulliken charges (electron), effective valence (electron) and Hirshfeld charge (electron) in $Mo_5PB_2$.

| Compound | Charge spilling | Species | Mulliken atomic populations | | | | Mulliken charge | Formal ionic charge | Effective valence | Hirshfeld charge | Effective valence |
|---|---|---|---|---|---|---|---|---|---|---|---|
| | | | s | p | d | Total | | | | | |
| $Mo_5PB_2$ | 0.22 | Mo | 2.25 | 6.86 | 4.98 | 14.08 | -0.08 | +4 | 3.92 | 0.12 | 3.88 |
| | | | 2.19 | 6.54 | 5.04 | 13.78 | 0.22 | +4 | 3.78 | 0.16 | 3.84 |
| | | P | 1.50 | 3.38 | 0.00 | 4.88 | 0.12 | -3 | 2.88 | -0.20 | 2.80 |
| | | B | 1.01 | 2.46 | 0.00 | 3.47 | -0.47 | -3 | 2.53 | -0.27 | 2.73 |

The calculated bond overlap population, bond length, metallic population and total number of each type of bond in $Mo_5PB_2$ are listed in Table 11. Materials with higher bond population, electron density, and smaller bond length exhibit high crystal stiffness and hardness Mulliken bond population analysis yields useful information regarding the bond overlap in compounds.

The degree of overlap of the electron clouds between two bonding atoms of a crystal is defined by the Mulliken bond populations. Bond order defines the overlap population of electrons between atoms. It also measures the strength of the covalent bond between atoms and the strength of the bond per the unit volume. The value of overlap population higher than zero and zero/close to zero indicates the presence of covalent and ionic bonds, respectively. The overlap population value close to zero implies that there is no significant interaction between the electronic populations of the two bonding atoms. In fact there are both positive and negative values of overlap populations found for $Mo_5PB_2$. The positive (+) and negative (-) values of bond overlap populations of a material point towards the states of bonding or antibonding nature of interactions between the atoms involved, respectively [96]. Thus, the bond overlap population values of $Mo_5PB_2$ indicate the presence of both bonding-type and anti-bonding-type interactions. Table 11 shows that both ionic and covalent bondings are present. In $Mo_5PB_2$, there are fourteen bonds which exhibit antibonding nature.

**Table 11**

The calculated Mulliken bond overlap population of $\mu$-type bond $P^\mu$, bond length $d^\mu$(Å), metallic population $P^{\mu'}$ and total number of bond $N^\mu$, of $Mo_5PB_2$.

| Compound | Bond | $P^\mu$ | $d^\mu$ | $P^{\mu'}$ | $N^\mu$ |
|---|---|---|---|---|---|
| $Mo_5PB_2$ | B-B | 0.43 | 1.99 | 0.023 | 4 |
| | | -0.05 | 3.38 | 0.023 | 8 |
| | | 0.00 | 5.73 | 0.023 | 8 |
| | B-Mo | 0.18 | 2.33 | 0.023 | 32 |
| | | 0.26 | 2.39 | 0.023 | 16 |
| | | 0.25 | 2.40 | 0.023 | 16 |
| | | -0.03 | 4.02 | 0.023 | 32 |
| | | -0.01 | 4.66 | 0.023 | 16 |
| | | 0.00 | 4.99 | 0.023 | 32 |
| | B-P | -0.02 | 3.66 | 0.023 | 32 |
| | P-Mo | 0.20 | 2.54 | 0.023 | 32 |
| | | 0.15 | 2.78 | 0.023 | 8 |
| | | 0.01 | 4.87 | 0.023 | 32 |
| | | -0.03 | 5.06 | 0.023 | 8 |

| | | | | |
|---|---|---|---|---|
| Mo-Mo | -0.29 | 2.72 | 0.023 | 32 |
| | -0.06 | 2.82 | 0.023 | 8 |
| | -0.69 | 2.82 | 0.023 | 8 |
| | -0.18 | 3.11 | 0.023 | 8 |
| | -0.11 | 3.15 | 0.023 | 16 |
| | -0.32 | 3.15 | 0.023 | 16 |
| | -0.03 | 4.20 | 0.023 | 8 |
| | -1.25 | 4.23 | 0.023 | 2 |
| | -0.09 | 4.43 | 0.023 | 16 |
| | -0.24 | 4.57 | 0.023 | 32 |
| | 0.51 | 4.88 | 0.023 | 8 |
| | 1.01 | 5.55 | 0.023 | 2 |
| | 0.01 | 5.73 | 0.023 | 8 |
| | -0.02 | 5.90 | 0.023 | 16 |
| P-P | 0.00 | 4.23 | 0.023 | 2 |
| | -0.01 | 5.55 | 0.023 | 2 |

## *3.8 Optical properties*

The optical response of a material is determined by a set of frequency (or equivalently, energy) dependent parameters that depend strongly on the details of the electronic band structure. The photons in the light incident on the surface of a material interact with the charge carriers and give rise to the optical characteristics. In recent times, interest in studying optical properties is increasing because of possible applications in integrated optics such as optical modulation, optoelectronics, optical information sector, and optical data storage. Study optical properties is also an essential tool for understanding other properties, like the electronic energy band structure, lattice vibrations, effect impurity levels, excitons, localized defects, and certain magnetic excitations [97].

Since the elastic properties of $Mo_5PB_2$ are directional dependent, optical parameters might also show anisotropic nature. Therefore, we have calculated optical properties of $Mo_5PB_2$ for two different polarization directions [100] and [001] of the incident electric field. The electronic band structures and density of states predict that $Mo_5PB_2$ is a metal. For metallic compounds, it is required to include Drude damping [98, 99] for the analysis of optical constants. The calculations of the optical constants of $Mo_5PB_2$ have been done using a screened plasma energy of 0 eV and a Drude damping of 0.05 eV as prescribed in the CASTEP code.

The optical properties of $Mo_5PB_2$ was calculated via the frequency dependent dielectric function,

$$\varepsilon(\omega) = \varepsilon_1(\omega) + i\varepsilon_2(\omega) \qquad (37)$$

where $\varepsilon_1(\omega)$ and $\varepsilon_2(\omega)$ are real and imaginary parts of the dielectric function, $\varepsilon(\omega)$. The real and imaginary parts of the dielectric function are associated with the group velocity of incident electromagnetic wave in the material and the absorption of energy within the material from the incident electromagnetic field due to the dipole motion, respectively. Intraband and interband charge transitions are two contributions to dielectric function in the condensed matter system. Intraband transitions are important at low energy, whereas the interband term is strongly controlled by the electronic band structure [100]. From complex dielectric function of a material, one can obtain the other energy (frequency) dependent optical constants, such as the refractive index, $n(\omega)$, extinction coefficient, $k(\omega)$, optical reflectivity, $R(\omega)$, absorption coefficient, $\alpha(\omega)$, energy-loss function, $L(\omega)$, and optical conductivity, $\sigma(\omega)$. The estimated optical constants of $Mo_5PB_2$ are shown in Figure 6, for the photon energy range up to 30 eV for [001] and [100] electric field polarization directions.

Figure 6(a) illustrates the variation of real and imaginary parts of dielectric function of $Mo_5PB_2$. It is observed from Fig. 6(a) that the value of $\varepsilon_2$ become zero at ~25.8 eV predicting that the material will become transparent above 25.8 eV. Generally, $\varepsilon_2$ becomes nonzero when the absorption occurs [35]. Moreover, Fig. 6(a) shows that the effective plasma frequency, $\omega_p$, is located at ~25.8 eV, for both the directions of electric field polarization. The real part of the dielectric function shows conventional metallic behavior with a strong Drude peak at low energy (Fig. 6(a)).

The complex refractive index of $Mo_5PB_2$ is expressed as: $N(\omega) = n(\omega) + ik(\omega)$, where $n(\omega)$ is the real part and $k(\omega)$ is termed as the extinction coefficient. The real part of the refractive index determines the group velocity of the electromagnetic wave inside a material, while the imaginary part (known as extinction coefficient) determines the amount of attenuation of the incident electromagnetic wave when it passes through the material. Since refractive index of a material is closely related to the electronic polarizability of ions and the local field inside the material, it is considered as one of the fundamental optical parameter. The complex refractive index spectra are shown in Fig. 6(b). The study of extinction coefficient, $k(\omega)$, of a material is crucial for designing photoelectric device. Figure 6(b) shows that the maximum value of n is at zero energy and decreases with increasing photon energy. The extinction coefficient exhibits strong peaks at ~1.5 eV for both polarization directions as displayed in Fig. 6(b).

The optical conductivity of a material can be defined as the conductivity of free charge carriers over a defined range of the photon energies. Figure 6(c) shows the calculated frequency-dependent optical conductivity $\sigma(\omega)$ of $Mo_5PB_2$ as a function of photon energy. The photoconductivity of $Mo_5PB_2$ starts from zero photon energy for both polarization directions, which implies that the materials have no band gap agreeing with the electronic band structure and TDOS calculations (Figs. 3). The photoconductivity of both directions increases with photon energy, reaches maximum (at ~0.89 eV and ~1.07 eV for [100] and [001] polarization directions, respectively), decreases gradually with further increase in energy and tends to zero at around 26.0 eV.

Mo$_5$PB$_2$ has high optical conductivity around 1.0 eV. Optical conductivity is much higher for [100] polarization compared to that for [001] polarization indicating an anisotropy in photo conductivity. The optical conductivity spectra are governed largely by the imaginary part of the dielectric function. This is seen from the qualitative agreement in the response spectra of σ(ω) and ε$_2$(ω) shown in Figs. 6(c) and 6(a), respectively.

Fig. 6(d) displays the reflectivity spectra of Mo$_5$PB$_2$ as a function of photon energy. The reflectivity is very high in the most of the visible spectra for both polarization directions. R(ω) decreases sharply at ~2.5 eV and levels off from about 4.0 eV. It is interesting to note that over a very wide spectral range from 4.0 eV to ~22.0 eV, R(ω) is nonselective. In this wide band of photon energies, the value of R(ω) remains above 45%. This implies that Mo$_5$PB$_2$ has potential to be used as a radiation reflector.

The absorption coefficient is an important parameter to understand electronic nature of a material; whether it is metallic, semiconducting or insulating. Optical absorption spectra is also helpful to determine the kind of electronic transition (indirect or direct transition or both) occurring in the energy bands [101]. The optical conductivity of a material increases as a result of absorption of photons. Figure 6(e) shows the absorption coefficient spectra of Mo$_5$PB$_2$ for [100] and [001] polarizations. The optical absorption of Mo$_5$PB$_2$ starts from zero photon energy consistent with the photoconductivity, dielectric function and band structure calculations. It is seen from Fig. 6(e) that absorption coefficient of Mo$_5$PB$_2$ is quite high in the region from ~6.5 eV to 23 eV. The position of the peak value of absorption coefficient for [001] polarization is different from that for [100] and it is higher along the [100] polarization direction compared to [001]. The peak value for [001] is around 15.71 eV, whereas for [100] the peak value is around 10.98 eV. For both polarizations, α(ω) decreases sharply at ~24.0 eV which in agreement with the position of the loss peak (Fig. 6(f)).

Figure 6(f) shows the electron energy loss function as a function of frequency L(ω) for Mo$_5$PB$_2$. A material's loss function is related to its absorption and reflection characteristics. In addition, a material's peak in the loss function L(ω) is also correspond to its trailing edges in the reflection spectra [102, 103], for instance, the peak in L(ω) is located at about ~25.0 eV corresponding to the abrupt reduction of R(ω). The loss function of a material becomes prominent at the onset of ε$_2$ < 1 and ε$_1$ = 0 [76, 104, 105] at high photon energy. The peak in L(ω) spectra represents the plasma resonance due to collective charge excitation and the corresponding frequency is the so-called plasma frequency, ω$_P$ [105]. The study of energy loss function is useful for understanding the screened charge excitation spectra, specially the collective excitations generated by a swift electron traversing a solid. It is observed that for [100] and [001] polarizations, the peaks of L(ω) are located at 25.9 eV and 25.5 eV, respectively. The sharp loss peak also represents the abrupt reduction in absorption coefficient of Mo$_5$PB$_2$ (Fig. 6(e)). At energies above the plasma resonance, Mo$_5$PB$_2$ becomes transparent to incident photons and exhibits optical features of an insulator.

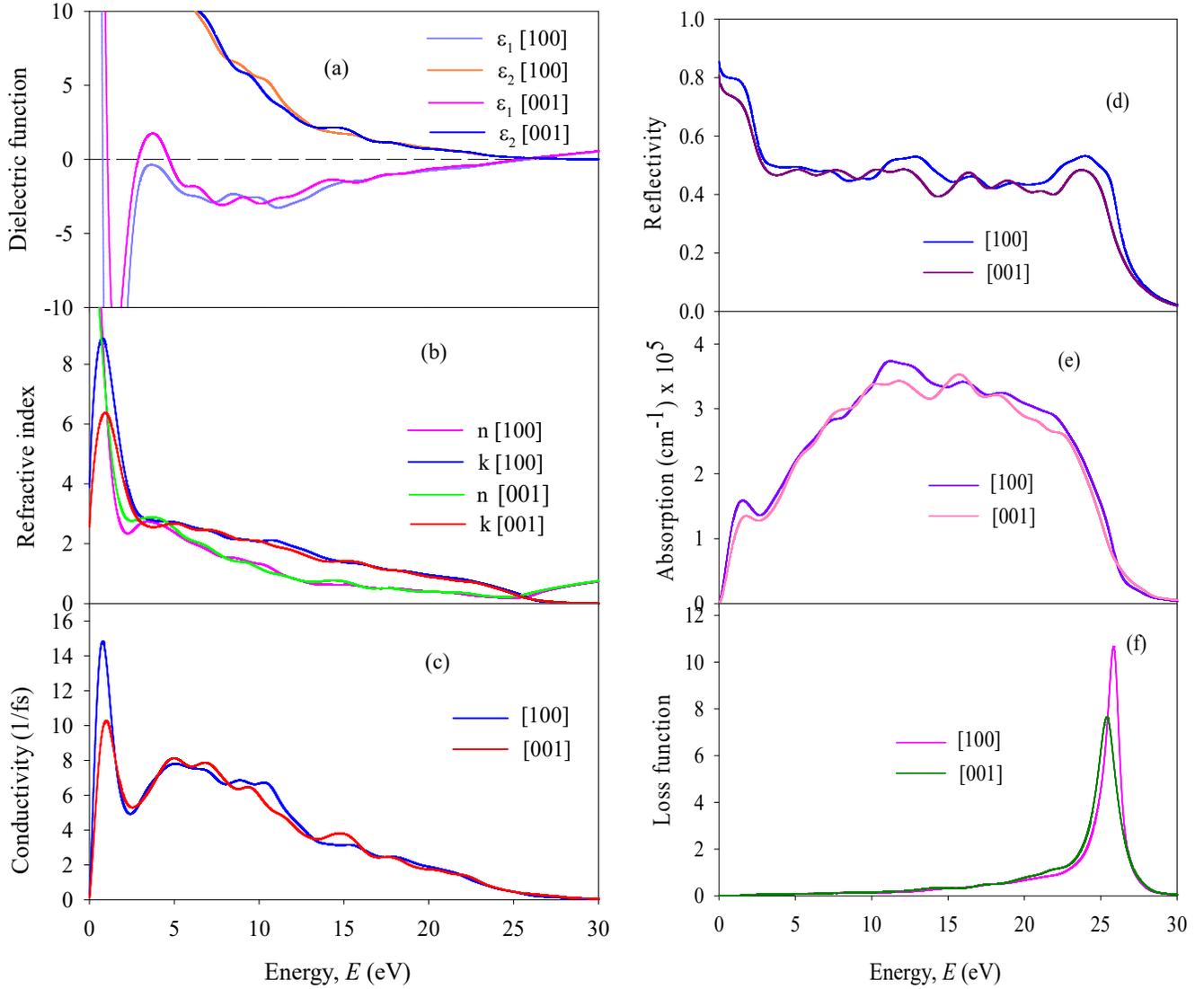

**Figure 6.** The photon energy dependent optical parameters of $Mo_5PB_2$ for different directions of electric field polarization. (a) Real and imaginary parts of dielectric function, (b) refractive index, (c) optical conductivity, (d) reflectivity, (e) absorption coefficient, and (f) loss function.

## 4. Discussion and conclusions

In this study we have investigated a large number of hitherto unexplored properties of $Mo_5PB_2$ in details, such as elastic, electronic, thermophysical, bonding and optoelectronic properties. The compound is found to be mechanically stable. $Mo_5PB_2$ possesses significant elastic anisotropy. The compound is machinable, ductile and possesses significant hardness. The bonding character is mixed with notable ionic and metallic contributions with some role played by covalent bondings. The electronic band structure calculations reveal metallic characteristics with a large

TDOS at the Fermi level. The calculated value of the Coulomb pseudopotential is also large. It is interesting to note that large TDOS at Fermi level enhances superconducting transition temperature, but the repulsive Coulomb pseudopotential hinders Cooper pairing [66, 106, 107]. The main contribution to the $N(E_F)$ comes from the Mo 4$d$ electronic states. We, therefore, predict that the major part of the superconducting condensation energy in $Mo_5PB_2$ comes from pairing of Mo 4$d$ electrons. Both the charge density distribution and electron density difference show a moderate level of direction dependency. The value of the universal log-Euclidean index ($A^L$) implies that the bonding strength in $Mo_5PB_2$ is anisotropic along different directions within the crystal and its comparatively lower value predicts the absence of significant layered characteristics.

The calculated value of Debye temperature is found to be in good agreement with the experimental value [12]. The estimated values of elastic moduli, Debye temperature, minimum thermal conductivity, melting temperature and thermal expansion coefficient imply that $Mo_5PB_2$ has significant promise to be used as a thermal barrier coating (TBC) material [35].

The optical parameters have been explored in details for the first time. $Mo_5PB_2$ possesses high reflectivity and absorption coefficient over extended range of energy. The material also has a large static refractive index. All these characteristics auger well for its use in the optoelectronic device sector. The optical anisotropy of the compound is found to be quite low. The optical constant spectra agree very well with the calculated electronic band structure.

In summary, we have investigated the elastic, mechanical, bonding, acoustic, thermal, bulk electronic (band, density of states, charge density distribution, electron density difference) and optoelectronic properties of $Mo_5PB_2$ in details in this paper, for the first time. The compound possesses several attractive mechanical, thermal and optoelectronic features which will be suitable for engineering and device applications. We anticipate that the results presented in this paper will stimulate researchers to investigate $Mo_5PB_2$, both theoretically and experimentally, in greater details in future.

**Acknowledgements**
S.H.N. acknowledges the research grant (1151/5/52/RU/Science-07/19-20) from the Faculty of Science, University of Rajshahi, Bangladesh, which partly supported this work.

**Data availability**
The data sets generated and/or analyzed in this study are available from the corresponding author on reasonable request.

**Author Contributions**

M.I.N. performed the theoretical calculations, contributed to the analysis and wrote the draft manuscript. M.A.A. performed theoretical calculations and contributed to the analysis. S.H.N. designed and supervised the project, analyzed the results and finalized the manuscript. M.I.N. and M.A.A. contributed equally in the manuscript. All the authors reviewed the manuscript.

**Additional Information**
**Competing Interests**
The authors declare no competing interests.